\documentclass[aps,pra,preprint]{revtex4-1}
\usepackage{graphicx}
\usepackage{amssymb}
\usepackage{amsmath}
\usepackage{color}
\usepackage{enumerate}
\usepackage{bm}
\begin{document}
\title{Dipolar Interaction and Sample Shape Effects on the Hysteresis Properties of 2$d$ Array of Magnetic Nanoparticles}
\author{Manish Anand}
\email{itsanand121@gmail.com}
\affiliation{Department of Physics, Bihar National College, Patna University, Patna-800004, India.}

\date{\today}
\begin{abstract}
We study the ground state and magnetic hysteresis properties of 2$d$ arrays ($L^{}_x\times L^{}_y$) of dipolar interacting magnetic nanoparticles (MNPs) by performing micromagnetic simulations. Our primary interest is to understand the effect of sample shape, $\Theta$- the ratio of the dipolar strength to the anisotropy strength, and the direction of the applied field $\vec{H} = H_{o}\hat{e}^{}_H$ on the ground state and the magnetic hysteresis in an array of MNPs. To study the effect of shape of the sample, we have varied the aspect ratio $A^{}_r=L^{}_y/L^{}_x$ which in turn, is found to induce shape anisotropy in the system. Our main observations are: (a) When the dipolar interaction is strong $(\Theta>1)$, the ground state morphology has in-plane ordering of magnetic moments. (b) The ground state morphology has randomly oriented magnetic moments which is robust with respect to system sizes and $A^{}_r$ for weakly interacting MNPs ($\Theta<1$). (c) Micromagnetic simulations suggests that the dipolar interaction decreases the coercive field $H^{}_c$. (d) The remanence magnetization $M^{}_r$ is found to be strongly dependent not only on the strength of dipolar interaction but also on the shape of the sample. (e) Due to anisotropic nature of dipolar interaction, a strong effect of shape anisotropy is observed when the field is applied along longer axis of the sample. The dipolar interaction in such a case induces an effective ferromagnetic coupling when the aspect ratio is very large. These results are of vital importance in high-density recording systems, magneto-impedance sensors, etc.
% Our work presents well guidelines regarding how magnetic hysteresis properties may change in the dipolar interacting MNPs arrays.
 %Our results are relevant for the experimental samples, which can be divided into various categories depending on how they behave when an external magnetic field is applied to them. 
 \end{abstract}
\maketitle

\section{Introduction}
Magnetic nanoparticles (MNPs) arrays are of profound importance not only because of interesting physics but also due to their numerous technological applications~\cite{wu2016,ulbrich,hervault,bustamante,schutz,dejardin,capiod,puntes,jordanovic,zhong,cowburn,manish2020}. For instance, a two-dimensional array of MNPs is a suitable system for high-density digital storage and perpendicular recording media~\cite{pardavi,mohtasebzadeh,xue}. The magnetic properties in such a case depend strongly on the shape, size, geometry of the systems and the magnetic interactions. The primary magnetic interaction in such systems is dipole-dipole interactions. The dipolar interaction is long-ranged and favours ferromagnetic as well as antiferromagnetic coupling. As a result, unconventional morphologies are observed when the size of the system is comparable to the range of dipolar interactions. Due to anisotropic nature of dipolar interactions, there have also been observations of ferromagnetic, striped, checkerboard patterns, vortices, etc. depending on the geometry of the lattice~\cite{Low,Kbell,Edlund,anand2019,alivisatos,collier}. The magnetic hysteresis properties in such systems are found to be greatly influenced by the dipolar interactions~\cite{kechrakos2005,usov2017,branquinho2013,anand2020}

%Collective phenomena, like magnetic hysteresis, are based upon the properties of individual magnetic nanoparticles and dipolar interactions.

There are several studies in the literature which make an implicit and explicit reference to the ubiquitous dipolar interactions and anisotropy in ordered arrays of MNPs. We summarize below few of them which is relevant to our present work:
(a) Li {\it et al.} studied the ground state, magnetic specific and magnetic hysteresis properties for three types of closely spaced nanomagnet arrays using Monte Carlo simulation~\cite{li}. They observed vortex state due to the dipolar interactions in these arrays. For face-centred cubic nanomagnet arrays, a slight jump occurs in the hysteresis curve. (b) Using the micromagnetic simulation, Yang {\it et al.} studied the magnetic properties in two-dimensional arrays of MNPs~\cite{yang}. They found that coercivity increases with arrays disorder. (c) Using the Landau-Lifshitz-Gilbert approach, Morales-Meza {\it et al.} studied the  magnetization reversal in a two-dimensional array of MNPs~\cite{morales}. They showed that coercivity is reduced even if the particle position in the array is random. (d) Chinni {\it et al.} performed experiments and micromagnetic simulation using Object Oriented MicroMagnetic Framework (OOMMF) to study magnetic properties in nanocomposite films~\cite{chinni}. They  observed an unexpected hysteretic behaviour characterized by in-plane anisotropy and crossed branches in the hysteresis curves measured along the hard  direction. (e) Faure {\it et al.} studied the magnetic properties of ordered arrays of MNPs using experiment and Monte Carlo simulation~\cite{faure}. The dipolar interaction is found to induce a ferromagnetic coupling that increases in strength with decreasing thickness of the array. (f) Using theoretical calculations, Xue {\it et al.} studied the magnetic hysteresis properties of the two-dimensional hexagonal array of MNPs~\cite{xue}. The hysteresis curves are found to vary their shapes  from a rectangle to a non-hysteresis straight line through a set of complicated loops, in accordance with the magnetization reversal process.
 
 Although the above studies imply that dipolar interaction and geometrical arrangement of MNPs affect the magnetic properties of an ordered array of MNPs, a systematic study as a function of dipolar interaction strength, shape anisotropy of the system and direction of the applied field is still missing. In this work, we attempt to understand the effect of dipolar interaction manipulated by changing the interparticle separation, the direction of applied magnetic field and shape anisotropy induced by varying the aspect ratio on the magnetic properties of two dimensional ($2d$) arrays of MNPs. The main questions which we have attempted to address are : (1) What are the consequences of dipolar interaction on ground state organizations of MNPs? (2) How are the magnetic hysteresis properties modified due to the dipolar interactions, shape anisotropy of the sample and direction of the applied magnetic field? To answer these questions, we consider 2$d$ ($L^{}_x\times L^{}_y$) arrays of cubical shaped Fe$_3$O$_4$ MNPs. It is well known fact that uniformly magnetized cubical MNPs do not possess shape anisotropy~\cite{moskowitz}. But we have been able to induce it in the system by just varying the sample shape. It will be shown that the latter has a drastic effect on the hysteresis behaviour. These ordered arrays of MNPs have many interesting properties due to structural order, well defined interparticle interactions and geometry confinement~\cite{Held}. In such a case, MNPs are coated with an inorganic surfactant to prevent agglomeration during self-assembling. Due to this MNPs are at a well distance beyond the range of exchange interaction. As a consequence, the predominant interaction in such arrays is dipolar interaction \cite{Dkech}. The energy of such an assembly is, therefore, given by the sum of the anisotropy energy and the dipolar interaction energy. 

To vary the relative dipolar interaction strength with respect to anisotropy strength, we have defined a ratio $\Theta = D/KV$ where $D$ is the strength of dipolar interaction, $K$ is the anisotropy constant,  and $V$ is the volume of the nanoparticle. The value of $\Theta$ for various interparticle separation $a$ for Fe$_3$O$_4$ has been given in Table~\ref{table1}. We refer to $\Theta > 1$ as the strong dipolar interaction regime, $\Theta < 1$ as the weak dipolar interaction regime.
We have performed micromagnetic simulation using OOMMF code  from NIST~\cite{donahue}. In OOMMF code, the finite difference method is employed, which requires discretization of a chosen geometry over a grid of identical prism-cells, and the magnetization is supposed to be uniform in each cell. {\color{black} In the implementation of OOMMF, the continuous magnetic material is divided in discrete cubes (i.e. grid cells), which are in geometrical contact. This discretization scheme suits to a
continuous magnetic film rather than an assembly of non-touching nanoparticles. Due to this reason, this type
of modelling does not discriminate between the magnetic behavior of 2$d$ continuous films and 2$d$
nanoparticle arrays. To overcome this, we have put exchange interaction to zero to mimic the seperated MNPs arrays magnetically. The absence of exhange interaction is also needed for the MNPs where they are coated with a surfactant to avoid agglomeration. We have sucessfully implemented it in our earlier studies~\cite{anand2016,anand2018}}. We have used 
Landau-Lifshitz-Gilbert (LLG) equation which is used to describe the precessional motion of a moment in magnetic field at $T = 0$ K. We have solved the coupled equation of motion for the given lattice to obtain the minimum energy configurations. To study the hysteresis, we have applied a dc magnetic field $\mu^{}_o\vec{H} = H^{}_{o}\hat{e}^{}_H$, where $\hat{e}^{}_H = \hat{x}$, $\hat{y}$ and $\hat{z}$ is the direction of the applied field along $x$, $y$ and $z$-axis respectively. We also study the role of aspect ratio $A_r = L^{}_y/L^{}_x$ to induce shape anisotropy in the system and the direction of the field $\hat{e}^{}_H$. 
%The main results of our study are as follows:\\
%\begin{enumerate}[(a)]
%\item The interaction between the magnetic moments is negligible for $\Theta <1$. The ground state morphologies consist of randomly oriented magnetic moments irrespective of sample size and aspect ratio. 
%\item There is local order in the ground state morphologies for $\Theta >1$. The magnetic moments prefer to orient in the plane of the sample.
%\item There is a tendency to align ferromagnetically as aspect ratio $A^{}_r$ is increased for strongly interacting MNPs ($\Theta >1$).
%\item  The dipolar interaction diminishes the coercive field $H_c$.
%\item Due to an increase in the shape anisotropy, $H^{}_c$ decreases as $A_r$ increases for $\Theta>1$.
%\item A strong effect of shape anisotropy is observed when the applied field is along the longer axis of the sample for $\Theta>1$. The dipolar interaction induces ferromagnetic coupling, which is reflected in the very large value of remanent magnetization $M^{}_r\approx0.9$.
%\item For strongly interacting MNPs, non-hysteresis is observed when the field is applied normal to the plane of the sample. 
%\end{enumerate}

The interplay of anisotropy, dipolar energy and aspect ratio creates unusual morphologies which have profound implications on the magnetic properties. Our work, therefore, throws light on this subject from a microscopic picture and provides a basis for results obtained in experimental and theoretical studies. This paper is organized as follows. In Section~II, we introduce the model for an array of MNPs, the LLG equation, which provides the prototypical ground state (GS) morphologies and equilibrium morphologies. In Sec.~III, we present our numerical results and discuss the dependence of the magnetic hysteresis  on $\Theta$, aspect ratio $A_r = L_y/L_x$  and the direction of the applied field $\hat{e}^{}_H$. A  conclusion of our results is provided in Section IV.

\section{Model and Methodology}
\subsection{Model for Dipolar Interacting Arrays of MNPs}
We consider a self-assembled 2$d$ array ($L_x\times L_y$) of cubical shaped magnetic nanoparticles in the $xy$-plane. The total number of the MNPs in the assmebly are $N = L_x/a\times L_y/a$, where $l$ is the length of an edge of the particle and $a$ is the lattice spacing.
%We consider a self-assembled 2$d$ arrays of cubical shaped MNPs in the $xy$-plane with dimensions $L_x\times L_y$. If $l$ is the length of an edge of the particle and $a$ is the lattice spacing. The total numbers of MNPs in the assembly are then $N = L_x/a\times L_y/a$. 
Each MNP has magnetic moment $\vec{\mu_i} = \mu \hat{e_i}$, $i = 1,2,\cdot\cdot\cdot,N$. The magnitude of the magnetic moment $\mu = M_sV$ where $M_s$ is the saturation magnetization and $V = l^3$ is the volume of the magnetic nanoparticle. The MNPs are assumed to have uniaxial anisotropy $\vec{K} = K \hat{k_i}$ where $K$ is the anisotropy constant, and $\hat{k_i}$ is the direction of the anisotropy. Generally, MNPs are coated with a surfactant to prevent agglomeration, which suppresses the exchange interactions. The energy of such a system, therefore, includes constribution from anisotroy energy $E^{}_a$ and dipolar interaction $E^{}_d$ \cite{Bedanta,Haase}:
%The energy of such an array, therefore, includes contributions from the magnetocrystalline anisotropy $E^{}_a$ and the dipole-dipole interactions $E^{}_d$ \cite{Bedanta,Haase}:
\begin{equation}
\label{ET1}
E = E_a + E_d = -KV\sum_{i}(\hat{k_i}\cdot{\hat{e}_i})^2 
 - D \sum_{i,j}\frac{3(\hat{e_{i}}\cdot {{\hat {r}}_{ij}})(\hat{e_{j}}\cdot {\hat {r}_{ij}}) - (\hat{e_{i}}\cdot\hat{e_{j}})}{\left(r_{ij}/a\right)^{3}} ,
\end{equation}
where $\mu_{o}$ is the permeability of free space, $\hat{e}_i$ is the unit vector along the magnetic moment, $r_{ij}$ is the distance between particles $i$ and $j$, and $\hat{r}_{ij}$ is the unit vector along with it. The $1/r_{ij}^3$ dependance means that the dipolar interaction is long ranged in nature. The analytical calculation of dipolar interaction between cubical shaped MNPs can be found in the reference~\cite{schabes}. It is well known fact that self assembled MNPs usually form $2d$ close-packed hexagonal lattices but magnetic properties are found to be nearly independent of the structure of the lattice~\cite{russier}. 

Here we define the {\it dipolar interaction strength} $D = \mu_{o} {\mu}^{2}/4\pi a^{3}$. As magnetic properties of such an assembly are governed by the relative strength of the anisotropy and dipolar energy, we have defined a ratio $\Theta = D/KV$. Although dipolar interactions and anisotropy energy depend on various system parameters, the behaviour of the assembly will be dictated by $\Theta$ rather than the precise values of parameters such as $a$, $V,$ $\mu$ and $K$. When $D$ is greater than $KV$, the dipolar interaction is stronger than anisotropy i.e $\Theta>1$. Similarly, $\Theta <1$ can be termed as the weak dipolar regime. 

In the presence of an external magnetic field $\mu^{}_{o}\vec{H}$, there is an additional contribution to energy $E$ given by \cite{Haase}
\begin{equation}
\label{EZ}
E_{H} = -  H^{}_{o}\sum_{i=1}\vec{\mu_{i}}\cdot \hat{e}^{}_H,
\end{equation}
here $H_{o}$ is the magnitude of field and $\hat{e}^{}_H$ is the unit vector in the direction of applied field. 
\subsection{Landau-Lifshitz-Gilbert Equation}
The precessional motion of the magnetic moment $\vec\mu_i$ in a magnetic field can be  described by the LLG equation~\cite{Dantas}:
\begin{equation}
\label{LL}
\frac{\mbox{d} \vec{\mu_i}}{\mbox{dt}} = - \gamma \vec{\mu_i} \times \vec{H}_{i}^e - \lambda\vec{\mu_i} \times \left( \vec{\mu_i} \times \vec{H}_{i}^{e}\right),\quad i=1,2,\cdot\cdot\cdot,N,
\end{equation}
where $\gamma$ is the electron gyromagnetic ratio, $ \lambda = \gamma \alpha/M_s$ is a phenomenological dimensionless damping factor, and $\vec{H}_{i}^{e} = -\partial E_T/\partial \vec{\mu}_i$ is the effective field experienced by magnetic moment, where $E_T=E+E^{}_{H}$. While studying ground state properties, $E^{}_H$ is taken to be zero. The first term in Eq.~(\ref{LL}) takes care of the precession of $\vec{\mu}_i$ around $\vec{H}_{i}^{e}$. The second term is due to  a phenomenological dissipative motion: the magnetic moment $\vec{\mu}_i$ precesses around $\vec{H}_{i}^{e}$.
%, looses energy to the environment in accordance with the damping factor $\lambda$ and eventually aligns along the effective field.  
The solution of these coupled differential equations yields the GS configuration $\{\vec{\mu_{i}}\}$, of the assembly. 
%The Conjugate-Gradient (CG) approach is used obtained the minimum energy spin configurations, which is implemented in OOMMF~\cite{donahue}. In this procedure, the evolution to a minimum energy configurations proceeds by a sequence of line minimizations. Each line represents a one-dimensional subspace in the 3$N$ dimensional space of possible magnetizations configurations, where $N$ is the number of spins in the simulations. Once a minimum is found along a line, a new direction is chosen which is ideally orthogonal to all preceding directions but related to energy gradient taken with respect to magnetizations. We start with a random orientation of the magnetic moments and stop the CG simulations at convergence $|\vec{\mu}\times\vec{H}_{}^e\times{\vec{\mu}}|\approx10^{-6}$.

To perform micromagnetic simulation using OOMMF code, the entire system is discretized into cells where each has lateral dimension $l$ say. Each cell in such case represents a magnetic moment $\mu = M_sV$. The centre-to-centre separation between moments is, therefore, $l \equiv a^{}_0$. By our formulation in Eq.~(\ref{ET1}), the strength of the dipolar interaction can be manipulated by varying the centre-to-centre separation $a$ of the MNPs. However, with the protocol implemented in OOMMF, a change in the centre-to-centre separation from $a_0$ to $a_{\beta}$ say, changes the cell volume  from $V \equiv V_0 \ (= a_0^3)$ to $V_{\beta} \ (= a_{\beta}^{3})$. As a consequence, the magnetic moment gets altered to $\mu_{\beta} = M_sV_{\beta}$, which ultimately modifies the magnetic properties of the particles under study. This undesirable artefact in the simulation needs to be overcome. To overcome this artefact, we formulated a rescaling method for saturation magnetization:
\begin{equation}
\label{mscale}
M_{s}^{\beta} = M_s^0\frac{V_0}{V_{\beta}},
\end{equation}
where $M_s^0$ is the saturation magnetization for nano particle of volume $V_0 = a_0^3$. It is easy to see that now $\mu = M_s^0 V_0 = M_{s}^{\beta}V_{\beta}$ as desired. Corresponding changes need to be incorporated in other related variables of interest such as the coercive field $H_c$ and the anisotropy field $H_K = 2K/M_s$ \cite{carrey} for non-interacting or weakly interacting MNPs which play an important role in hysteresis:
\begin{eqnarray}
\label{fscale1}
H_c^{\beta} &=& H_c^{0} \frac{V_{\beta}}{V_{0}},\\
\label{fscale2}
H_K^{\beta} &=& H_K^{0} \frac{V_{\beta}}{V_{0}}.
\end{eqnarray}

We have successfully implemented  this scaling procedure to study the heat dissipation and spin transport properties in the assembly of dipolar interacting MNPs in our earlier work~\cite{anand2016,anand2018}.

\section{Numerical Results}
We consider cubical shaped MNPs of Fe$_3$O$_4$ arranged on a $2d$ ($L_x \times L_y$) lattice. The particle size is chosen to be $l =10$ nm. We have used anisotropy constant $K = 13\times 10^{3}$ J$\mathrm{m}^{-3}$ and saturation magnetization $M^{}_s=4.77\times 10^{5}$ A/m for numerical evaluations. The six values of interparticle separation $a$ are considered for magnetic hysteresis study: $a_0 = 10$ nm, $a_1$ = 12 nm, $a_2$ = 16 nm, $a_3$ = 20 nm, $a_4$ = 30 nm and $a_5$ = 40 nm. Table~\ref{table1} provides the values of $\Theta$ for these interparticle separations. The initial condition that we choose for the assembly of MNPs is random orientations of magnetic moments and anisotropy axes. All the data obtained using the simulations are averaged over 50 sets of initial conditions. The ground state morphologies are obtained by solving LLG equation using OOMMF code in the absence of external magnetic field. To study the magnetic hysteresis properties, we apply a dc magnetic field of strength -150 mT to 150 mT for $a=10$ nm. For other interparticle separations, the scaled saturation magnetization $M_{s}^{\beta}$ were obtained using Eq.~(\ref{mscale}). The corresponding $H_{c}^{\beta}$ and $H_{K}^{\beta}$ were calculated using Eqs.~(\ref{fscale1}) and (\ref{fscale2}), respectively. We have chosen the magnetic field to be large enough as comapred to $H_c$ to achieve saturation and the ramping up or slowing down has been appropriately fine-tuned to capture the magnetic properties near $H_c$. We study the dependence of magnetic hysteresis on the $\Theta$, aspect ratio $A_r = L_y/L_x$ and the direction of the applied field $\hat{e}_H$. In our simulations, $L_x = 120$ nm; $A_r = 1, 2, 4, 8, 16$ and 32; $\Theta=1.75$, 1.01, 0.42, 0.22, 0.06 and 0.03; $\hat{e}^{}_{H}=\hat{x}$, $\hat{y}$ and $\hat{z}$.
\subsection{{\bf Ground State (GS) Morphologies}}
\begin{figure}[!htb]
\centering\includegraphics[scale=0.50]{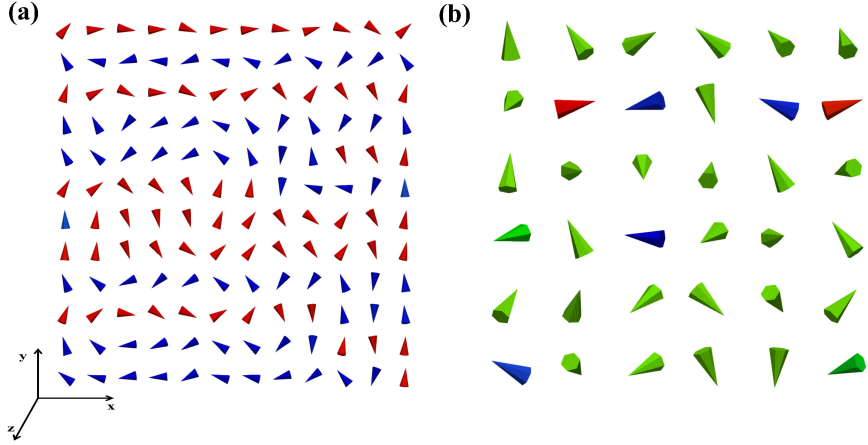}
\vspace{0.5cm}
\caption{Typical GS morphologies for: (a) $L_x = L_y = 120$ nm, $a=10$ nm and $\Theta = 1.75$; (b) $L_x = L_y = 120$ nm, $a=20$ nm and $\Theta = 0.22$. The magnetic moments are coloured with red colour cone, which has a positive $x$ component, and the magnetic moments with negative $x$ component are coloured blue colour cone. Those with a non-zero $z$ component, and pointing normal to the $xy$-plane, are indicated with shown with green colour cone. For $\Theta\simeq 0$, moments are randomly oriented, and they also tend to orient normal to the plane of the sample. For $\Theta>1$, the moments predominantly lie in the $xy$-plane and exhibit local order.} 
\label{figure1}
\end{figure}

\begin{figure}[!htb]
\centering\includegraphics[scale=0.50]{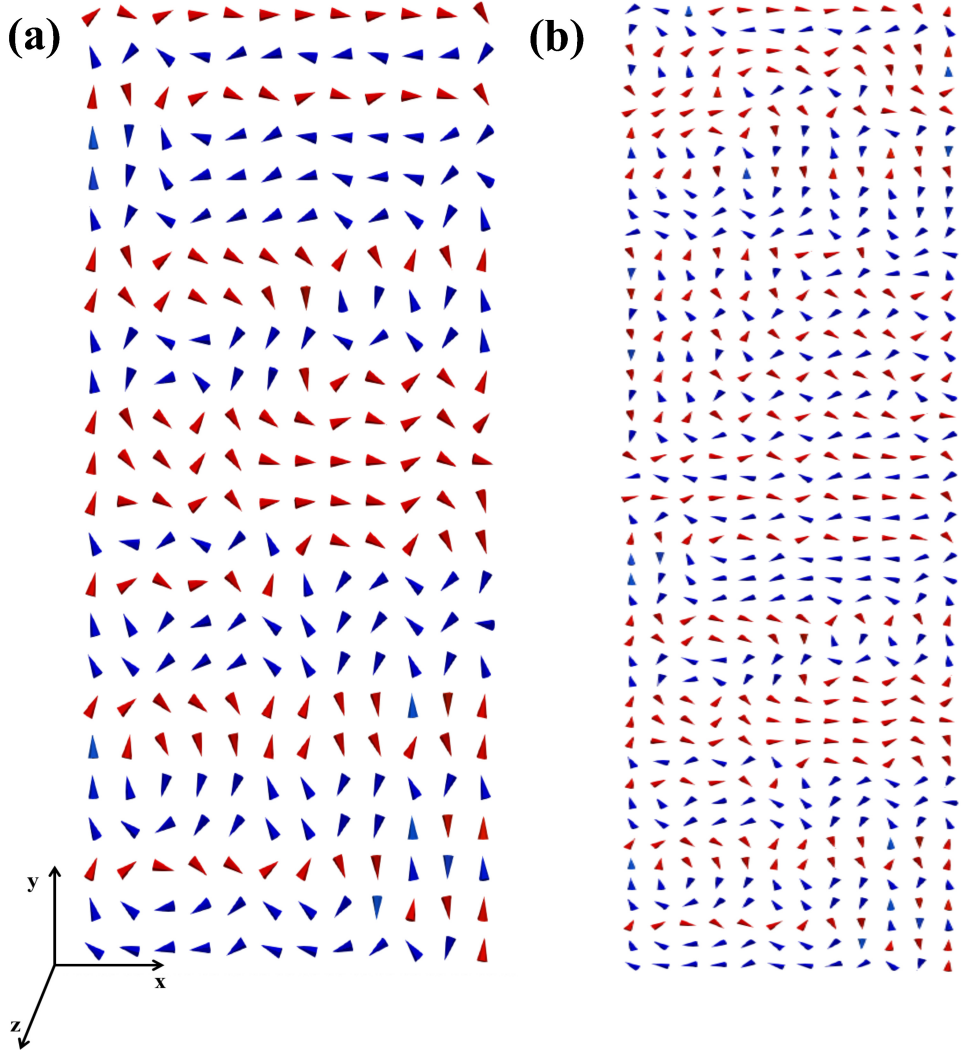}
\vspace{0.5cm}
\caption{Typical GS morphologies for strongly interacting MNPs ($a=10$ nm, $\Theta = 1.75$) as a function of aspect ratio $A^{}_r$. (a) $L_x =120$ nm and  $A_r = 2$ (b) $L^{}_x=120$ nm  and $A_r=4$. Due to large dipolar strength, there is an order in the morphology, and they also tend to orient in the plane of the sample. As $A^{}_r$ is increased, magnetic moments also favour chain arrangement along the longer the axis of the sample. The colour coding is the same as described in Fig.~(\ref{figure1}).}
\label{figure2}
\end{figure}

We first study the effect of $\Theta$ and $A^{}_r$ on ground state spin morphologies. For this purpose, we look square samples i.e $A_r = 1.0$ such that $L_x\times L_y \equiv L_x \times L_x$. Fig.~(\ref{figure1}) depicts GS morphologies for $L_x$ = 120 nm. The interparticle separation $a$ is chosen to be 10 nm [see Fig.~\ref{figure1}(a)] and 20 nm [see Fig.~\ref{figure1}(b)]. These correspond to $\Theta =  1.75$ and 0.22, respectively. So the number of spins in Fig.~\ref{figure1}(a) is  $12\times 12$ and $6\times 6$ in Fig.~\ref{figure1}(b). we depict the spins with noz-zero $z$-component by the green cone. Those lying in the $xy$ plane with a positive $x$-component have been indicated by the red coloured cone, while those with negative $x$-component by the blue coloured cone. It is quite evident that for strongly interacting MNPs ($\Theta = 1.75$), {\it all} the spins are in the $xy$-plane [see Fig.~\ref{figure1}(a)]. They exhibit locally ordered regions. On the other hand, when magnetic interactions are weak ($\Theta = 0.22$), the moments are randomly oriened, signifying lack of magnetic order [see Fig.~\ref{figure1}(b)]. Magnetic moments also tend to align normal to the plane of the sample. It means that dipolar interaction favours in-plane ordering. Then, we study the effect of aspect ratio $A_r$ on the spin morphologies. For $\Theta<1$, the features are unchanged as $A_r$ is increased to 2, 3, 4, etc. and are prototypically represented by Fig.~\ref{figure1}(b). We do not show them to avoid repetition. In Fig.~(\ref{figure2}), we depict morphologies corresponding to the strongly dipolar interacting MNPs ($a=10$ nm, $\Theta = 1.75$) for  $L_x$ = 120 nm and two aspect ratio (a)  $A_r$ = 2 [see Fig.~\ref{figure2}(a)] and (b) $A_r$ = 4 [see Fig.~\ref{figure2}(b)]. In both the cases, the MNPs exhibit local order and perfer to lie in the $xy$-plane. The morphology of the magnetic moments near the sample edges is very distinct from that in the bulk. The moments tend to align along the edges as ferromagnetic chains. 
\subsection{{\bf Magnetic Hysteresis Study}}
\begin{figure}[!htb]
\centering\includegraphics[scale=0.50]{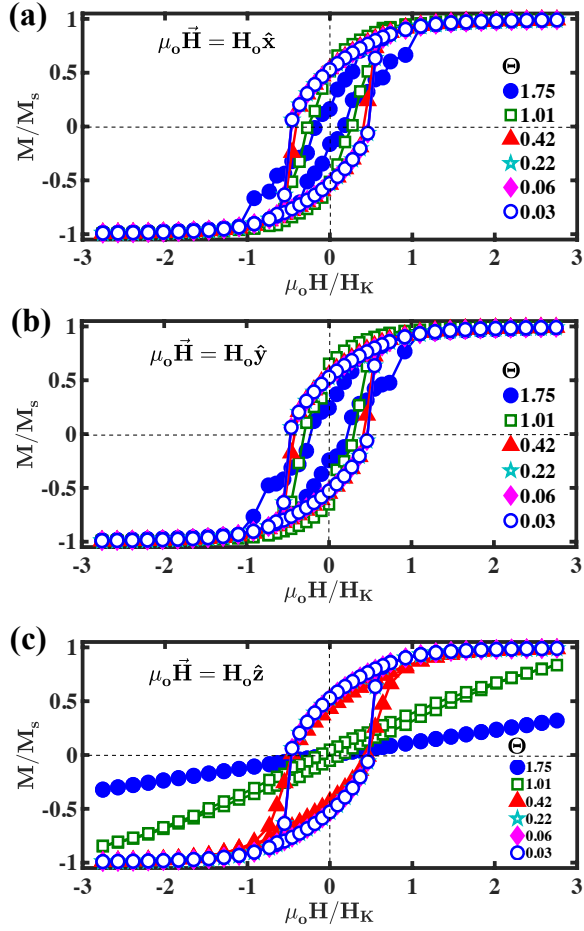}
\vspace{0.5cm}
\caption{Magnetic hysteresis behaviour as a function of dipolar interaction strength and direction of the applied field for square sample, i.e., $A^{}_r=1.0$ ($L^{}_x=L^{}_y=120$ nm). (a) The external field $\mu^{}_o \vec{H}$ is applied along the $x$-direction, (b) $\mu^{}_o \vec{H}=H_o\hat{y}$ and (c) $\mu^{}_o \vec{H}=H_o\hat{z}$. The strength of dipolar interaction has been varied from very large value $1.75$ to $0.03$. When the field is applied in the plane of the sample, coercive field $H^{}_c$ and remanent magnetization decrease with an increase in the strength of the dipolar interaction. When the field is applied is normal to the plane of the sample, $H^{}_c$ and $M^{}_r$ decrease to zero for large dipolar strength. The coercive field $H^{}_c\approx 0.48H^{}_K$ and $M^{}_r \approx 0.5$ are observed for weakly interacting MNPs irrespective of the direction of the applied field as expected (signature of the Stoner-Wohlfarth model).} 
\label{figure3}
\end{figure}

Next, we study the effect of dipolar strength $\Theta$, shape anisotropy of the sample and direction of applied field $\hat{e}^{}_H$ on the magnetic hysteresis in a systematic manner. To vary the dipolar interaction strength, we have varied the interparticle separation. To induce shape anisotropy in the system, we have changed the aspect ratio $A^{}_r$ of the sample. In Fig.~(\ref{figure3}), we have plotted magnetic hysteresis curves for square sample, i.e., $L_x=L_y=1.0$ as a function of $\Theta$ and $\hat{e}^{}_H$. We have considered six values of dipolar strength $\Theta= 1.75$, 1.01, 0.42, 0.22, 0.06 and 0.03 in each case. The magnetic field axis ($x$-axis) has been scaled by $H^{}_K$, the anisotropy field (for non-interacting MNPs with randomly oriented anisotropy axes). The direction of the external field is (a) $\mu_{o}\vec{H}=H^{}_{o}\hat{x}$ [see Fig.~\ref{figure3}(a)], (b) $\mu_{o}\vec{H}=H^{}_{o}\hat{y}$ [see Fig.~\ref{figure3}(b)] and (c) $\mu_{o}\vec{H}=H^{}_{o}\hat{z}$ [see Fig.~\ref{figure3}(c)]. It is quite evident that when the field is applied in the plane of the sample, i.e., either along $x$ or $y$-direction, the coercive field $H_c$ decreases with an increase in the strength of dipolar interaction. As a consequence, the area under the hysteresis curve diminishes. When the field is applied normal to the plane of the sample ($z$-direction), magnetic moment ceases to follow the applied field for $\Theta>1$. As a result, non-hysteresis is observed, so $H^{}_c$ and $M^{}_r$ tend to zero for strongly interacting MNPs. It is clearly seen that the coercive field $H^{}_c\approx 0.48 H^{}_K$ and $M^{}_r \approx 0.5$ for weak interaction or non-interacting MNPs ($\Theta<1$) irrespective of the direction of the applied field. These values of $H^{}_c$ and $M^{}_r$ correspond to single particle hysteresis with randomly oriented anisotropy axis~\cite{carrey,stoner1948}. It means that magnetic hysteresis follows the Stoner-Wohlfarth model when the interaction among the MNPs is weak as expected~\cite{carrey, stoner1948}.

\begin{figure}[!htb]
\centering\includegraphics[scale=0.40]{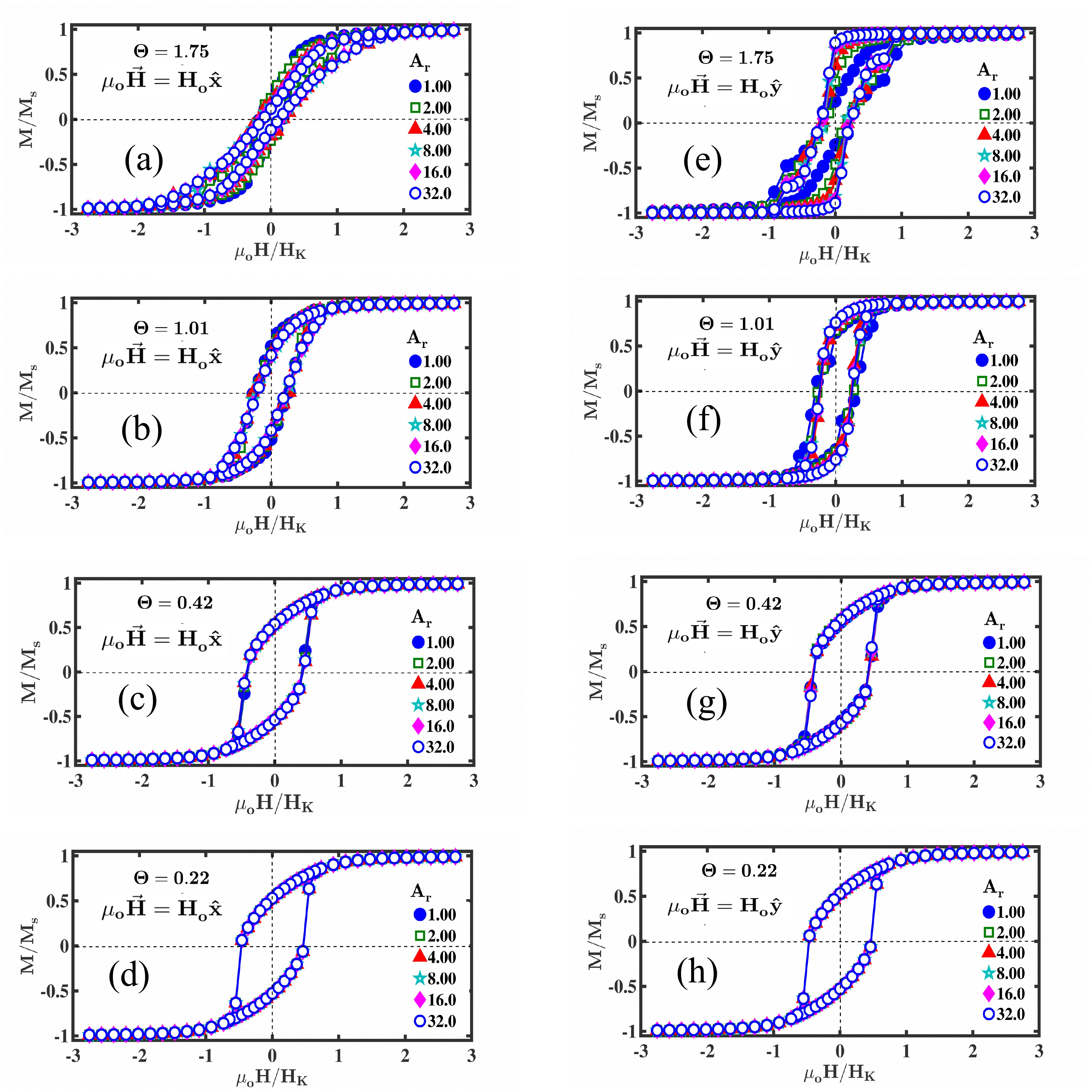}
%\vspace{0.5cm}
\caption{Dependence of magnetic hysteresis on aspect ratio $A^{}_r$ and dipolar interaction strength $\Theta$ when the field is applied in the plane of the sample the sample, i.e., along $x$-direction [Left Panel: Figure 4(a)-(d)] or $y$-direction [Right Panel: Figure 4(e)-(h)]. Four values of $\Theta =$ 1.75, 1.01, 0.42 and  0.22 have been considered. The aspect ratio $A^{}_r$ is varied from 1.0 to 32.0 in each case. The dipolar interaction decreases the coercive field $H_c$ and $M_r$.  There is a strong dependence of $M^{}_r$ on the shape anisotropy introduced by increasing $A^{}_r$,  $M^{}_r$ reaches to $0.9$ when the field is applied along the longer axis of the sample in the presence of strong dipolar interaction. For weak dipolar interaction, i.e., $\Theta  = 0.22$, magnetic hysteresis follow the Stoner-Wohlfarth model ($H^{}_c \approx 0.48 H^{}_K$ and $M_r \approx 0.5$) irrespective of $A^{}_r$.}
\label{figure4}
\end{figure}

\begin{figure}[!htb]
\centering\includegraphics[scale=0.40]{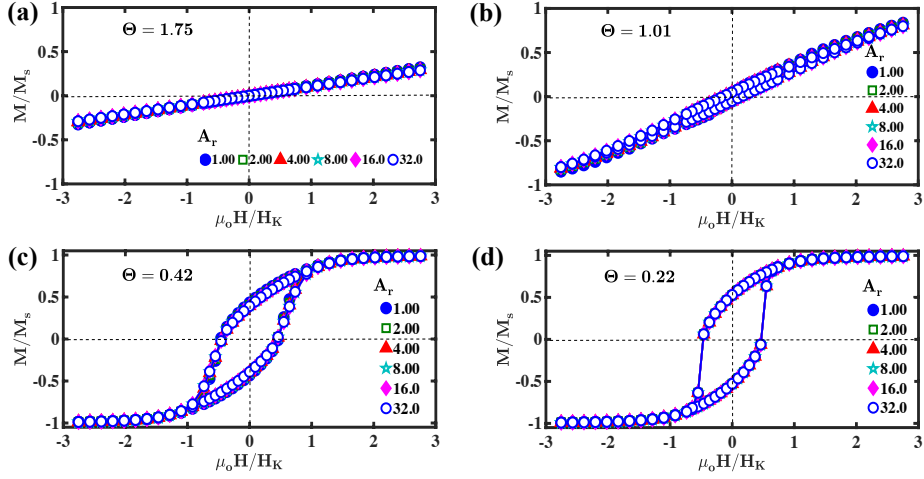}
%\vspace{0.5cm}
\caption{ Magnetic hysteresis as a function of dipolar interaction strength $\Theta$ and aspect ratio $A^{}_r$ when the field is applied normal to the plane of the sample $\mu_o\vec{H}=H_o\hat{z}$. Four values of $\Theta$=  1.75, 1.01, 0.42 and 0.22 have been considered. The aspect ratio $A^{}_r$ is varied from 1.0 to 32.0 in each case. For large  dipolar interaction strength, the magnetic moments cease to align normal to the sample as a result, non-hysteresis is observed. The coercive field $H^{}_c$ and $M^{}_r$ drop down to zero in this case.  For weak dipolar interaction, i.e., $\Theta  = 0.22$, magnetic hysteresis follow the Stoner-Wohlfarth model ($H^{}_c \approx 0.48 H^{}_K$ and $M_r \approx 0.5$) irrespective of the aspect ratio $A^{}_r$.}
\label{figure5}
\end{figure}

To study the dependence of shape anisotropy on the magnetic properties, we study the magnetic hysteresis as a function of aspect ratio $A^{}_r$ for various values of $\Theta$ and three directions of the applied field, i.e., along ${x}$, ${y}$ and ${z}$-axis respectively. We have considered six values of aspect ratio $A^{}_r =1.0$, 2.0, 4.0, 8.0, 16.0 and 32.0 in each case. In Fig.~(\ref{figure4}), the direction of the applied field is in the plane of the sample [$\mu_{o}\vec{H}=H_{o}\hat{x}$ and $\mu_{o}\vec{H}=H_{o}\hat{y}$]. The values of $\Theta$ are (a) $\Theta = 1.75$ [see Fig.~\ref{figure4}(a) and 4(e)], (b) $\Theta = 1.01$ [see Fig.~\ref{figure4}(b) and 4(f)], (a) $\Theta = 0.42$ [see Fig.~\ref{figure4}(c) and 4(g)] and (d) $\Theta = 0.22$ [see Fig.~\ref{figure4}(d) and 4(h)]. The direction of the applied field is normal to the plane of the sample (along the $z$-axis) in Fig.~(\ref{figure5}). The other parameters, $A^{}_r$ and $\Theta$ in Fig.~(\ref{figure5}) remain the same as that of Fig.~(\ref{figure4}). When the applied field is along the shorter axis (along the $x$-axis), dipolar interaction decreases the value of $H_c$ and $M_r$ [see Fig.~\ref{figure4}(a)-(d)]. There is a weak dependence of $H_c$ on $A_r$. For weakly interacting MNPs, hysteresis curves follow the Stoner-Wohlfarth model~\cite{stoner1948}. A strong effect of shape anisotropy is observed when the field is applied along the increasing length of the sample, i.e., along the $y$-axis.  The remanent magnetization $M^{}_r$ increases with increase in $A^{}_r$, and it reaches to 0.9 for strongly interacting MNPs [see Fig.~\ref{figure4}(e)]. In this case, also, the larger is the strength of dipolar interaction, the smaller is the value of $H^{}_c$ [see Fig.~\ref{figure4} (e)-(h)]. The magnetic moments cease to follow the applied magnetic field when the field is applied along the $z$-axis for strongly interacting MNPs. As a consequence, almost no hysteresis is observed for strongly interacting MNPs with applied field normal to the plane of the sample [see Fig.~\ref{figure5}(a)-(b)]. Like the other two cases stated above, weakly interacting MNPs follow the Stoner-Wohlfarth model irrespective of $A^{}_r$.
\subsection{{\bf Characterization of Magnetic Hysteresis Curves}}
\begin{figure}[!htb]
\centering\includegraphics[scale=0.40]{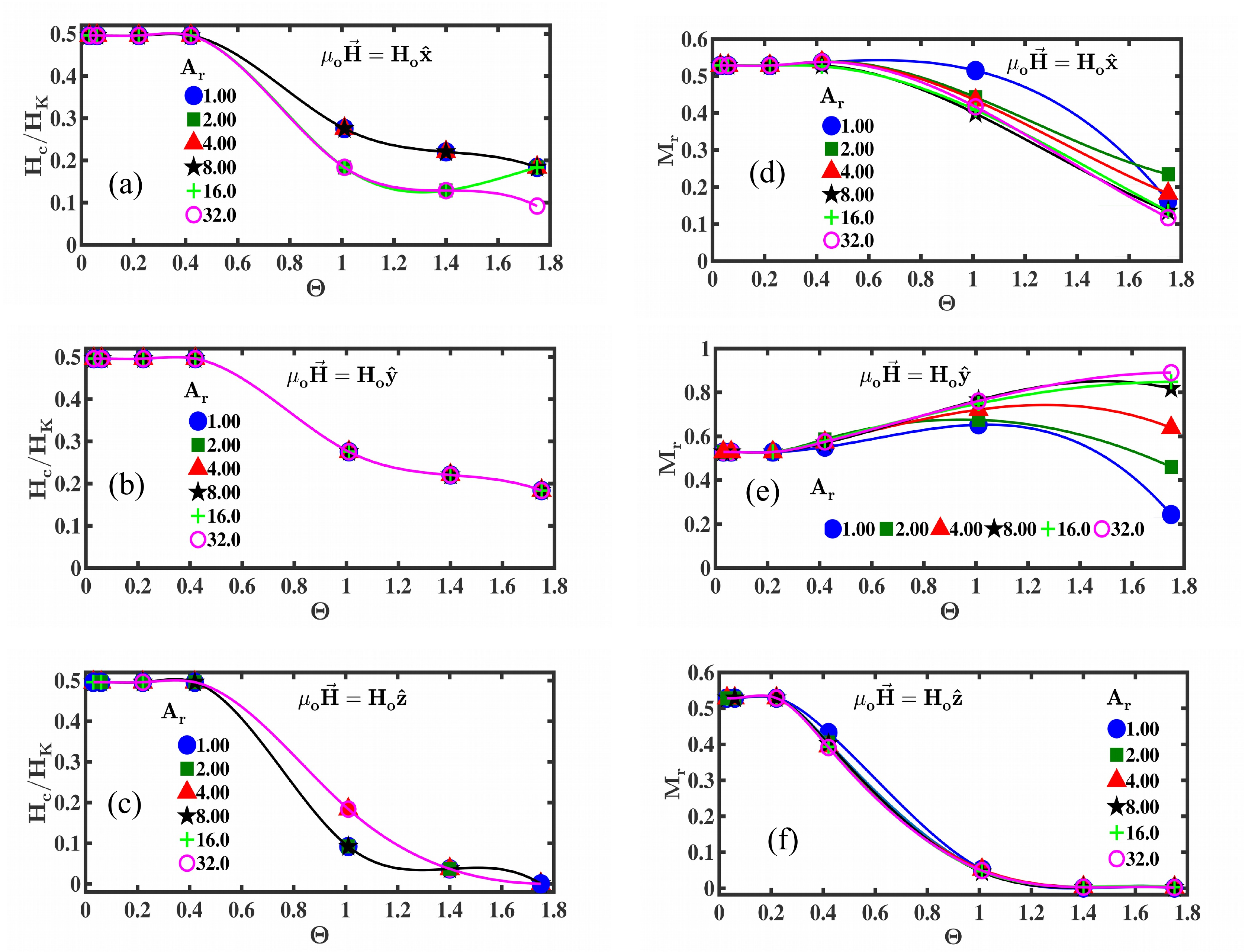}
\vspace{0.5cm}
\caption{Coercive field $H^{}_c$ and remanent magnetization $M^{}_r$ variation as a function of dipolar interaction strength $\Theta$, aspect ratio $A^{}_r$ and direction of the applied field. $H_c$ has been scaled by single particle anisotropy field $H^{}_K$. $H^{}_c$ decreases with an increase in dipolar strength $\Theta$ and $A^{}_r$ when the field is applied in the plane of the sample (either along $\hat{x}$ or $\hat{y}$-axis). $H^{}_c$ drops down to zero when the field is applied normal to the plane of the sample for strongly interacting MNPs. When the field is applied along the $x$-direction, $M^{}_r$ is decreased with an increase in aspect ratio and dipolar interaction strength $\Theta$.  There is a strong shape anisotropy effect is observed when the field is applied along the longer axis of the sample. $M^{}_r$ increases with increase in $\Theta$ and $A^{}_r$, and it reaches to 0.9 with $A^{}_r=32.0$ and $\Theta = 1.75$. When the external field is applied normal to the plane of the sample $\mu_o\vec{H}=H_o\hat{z}$, $M^{}_r$ decreases with increase in $A^{}_r$ and $\Theta$, and it drops down to zero for $\Theta = 1.75$ and $A^{}_r = 32.0$. }
\label{figure6}
\end{figure}

Finally, we study the variation of the coercive field $H^{}_c$ and $M^{}_{r}$ as a function of $\Theta$, $A^{}_r$ and $\hat{e}^{}_{H}$ by extracting their values from the simulated magnetic hysteresis curves.  In Fig.~(\ref{figure6}), we have plotted variation of $H^{}_c$ (scaled by $H^{}_K$) and $M^{}_r$ as function of dipolar strength $\Theta$ for six values of aspect ratio $A^{}_r=1.0$, 2.0, 4.0, 8.0, 16.0 and 32.0. The direction of the applied magnetic field is (a) $\mu_o\vec{H}=H_o\hat{x}$ [see Fig.~\ref{figure6}(a) and 6(d)], (b) $\mu_o\vec{H}=H_o\hat{y}$ [see Fig.~\ref{figure6}(b) and 6(e)] and (c) $\mu_o\vec{H}=H_o\hat{z}$ [see Fig.~\ref{figure6}(c) and 6(f)]. With the field applied along the $x$-axis, $H^{}_c$ decreases with an increase in the strength of dipolar interaction $\Theta$ and its value drops to $0.1H^{}_K$ for $A^{}_r=32.0$ and $\Theta = 1.75$ [see Fig.~\ref{figure6}(a)]. The variation of $H^{}_c$ is similar as with field along $x$-axis when the field is applied longer axis of the sample (along the $y$-axis), but the value of $H^{}_c$ is slightly higher in this case for $A_r=32.0$ and $\Theta=1.75$ [see Fig.~\ref{figure6}(b)]. When the field is applied normal to the plane of the sample, $H^{}_c$ decreases very fast with an increase in $\Theta$ and $A^{}_r$. Its value drops down to zero for strongly interacting MNPs, and $A_r=32.0$ [see Fig.~\ref{figure6}(c)]. There is a weak dependence of $M^{}_r$ on $A^{}_r$ when the field is applied along the $x$-axis [see Fig.~\ref{figure6}(d)]. There is a strong effect of shape anisotropy when the field is applied along $y$-axis [see Fig.~\ref{figure6}(e)]. For strongly interacting MNPs, $M^{}_r$ increases as the aspect ratio $A^{}_r$ is increased and it reaches to 0.9 for $\Theta=1.75$, and $A^{}_r=32.0$  [see Fig.~\ref{figure6}(e)]. $M^{}_r$ decreases very sharply with $\Theta$, and its value drops to zero for $\Theta=1.75$ when the magnetic field is applied normal to the plane of the sample [see Fig.~\ref{figure6}(f)]. For weakly interacting MNPs, magnetic hysteresis curves follow the Stoner-Wohlfarth model irrespective of aspect ratio and direction of the applied field, which is reflected in $H^{}_c$ ($\approx 0.48H^{}_K$) and $M^{}_r$ ($\approx 0.5$).

These results can be explained by probing the effect of dipolar interaction and shape anisotropy of the sample induced by varying aspect ratio. It is evident from the ground state morphologies that the dipolar interactions favour in-plane ordering of magnetic moments. When $A^{}_r$ is increased, the dipolar interaction favours ferromagnetic coupling between the magnetic moments. It is because of the fact the demagnetization field decreases as shape anisotropy is increased ($A^{}_r$ increases in our case), which is also reported by Wysin~\cite{wysin2012}. As a consequence, magnetic moments tend to align along the longer axis of the sample as it costs less energy as compared to other configurations. It has also been reported by jordanovic {\it et al.}~\cite{jordanovic}. This ferromagnetic coupling for strongly interacting MNPs can also be explained by that fact that the exchange interaction between the MNPs is assumed to be zero in the present work. Due to the absence of exchange interaction, the energetics of domain walls are entirely controlled by the dipolar interaction. It makes energetically favourable to form domain walls along the increasing length of the sample. When an external field is applied to these systems system, the response of this system not only depends on the dipolar strength but also on the shape anisotropy of the sample which has been induced by increasing $A^{}_r$ in our work. When the field is applied along the shorter axis of the sample, i.e., along the $x$-axis, the natural tendency to get aligned along the longer axis of the sample is hindered. As a result, magnetic moment ceases to follow the external field, which is reflected in a decrease in the value of $M^{}_r$ ($\approx 0.1$) and $H^{}_c$ to 0.1$H^{}_K$ for strongly interacting MNPs with $A^{}_r=32.0$. Magnetic moments are found to follow the field when the field is applied along the longer axis of the sample, but due to the increase in the strength of dipolar interaction as well, $H^{}_c$ decreases, but its value remains slightly higher as that of the previous situation. Due to the same reason, $M^{}_r$ increases with $A^{}_r$ for $\Theta >1$. This increase of $M^{}_r$ and decrease of $H^{}_c$ when the field is applied along the longer axis of the sample is in qualitative agreement with the work done by Garc{\'\i}a-Arribas {\it et al.}~\cite{garcia2013}. They have studied the shape anisotropy effect in a thin film permalloy microstrips. 

When we force the magnetic moments to get aligned normal to the plane of the sample by applying the field along the $z$-axis, they oppose strongly for a large value of $\Theta$ and $A^{}_r$. This happens due to the fact that the natural tendency of the magnetic moments is to get organized in the plane of the sample ($xy$-plane in our case). As a result, non-hysteresis is observed in this case,  which is reflected in zero value of $H^{}_c$ and $M^{}_r$.  This non-hysteresis behaviour has also been reported by Xue {\it et al.}~\cite{xue}. They have studied the hysteresis properties in a two-dimensional hexagonal array with aligned uniaxial anisotropy. From all the cases, it is quite evident that dipolar interaction always decreases the value of coercive field $H^{}_c$. It is because of that dipolar interactions cause a collective reversal of the magnetic moments under an applied magnetic field and as a consequence, the coercive field decreases with an increase in dipolar interaction strength.

\section{Conclusion}
To conclude, we have studied the effect of dipolar interactions and shape of the sample on the magnetic properties of 2$d$ ($L^{}_x\times L^{}_y$) array of cubical shaped magnetic nanoparticles (MNPs) by performing micromagnetic simulation using OOMMF code~\cite{donahue}. It is well known that uniformly magnetized cubical shaped  particles do not possesses shape anisotropy~\cite{moskowitz} but we have been able to induce the latter in the system by varying aspect ratio $A^{}_r$. Our primary aim was to study ground state (GS) morphologies, and understand the magnetic hysteresis properties as a function of $\Theta$, shape anisotropy induced by varying the aspect ratio $A_r (= L_y/L_x)$ and the direction of the applied field $\mu^{}_{o}\vec{H} = H^{}_{o}\hat{e}_H$. Our observations are as follows: (a) For  weakly interacting MNPs ($\Theta <1$), the magnetic moments are randomly oriented, and the morphology is unaffected by $A_r$. MNPs also tend to align normal to the plane of the sample. (b) For strong dipolar strengths ($\Theta >1$), magnetic moments prefer to orient in the plane of the sample. The morphology, in this case, comprises regions of correlated moments. (c) Magnetic moment favours ferromagnetic coupling along the longer axis of the sample when $A^{}_r$ is increased provided $\Theta>1$. (d) When the dipolar interaction is weak; magnetic hysteresis curves follow the Stoner-Wohlfarth model ($H^{}_c\approx 0.48H^{}_K$ and $M^{}_r \approx 0.5$) irrespective of the direction of applied field $\hat{e}^{}_H$ and aspect ratio $A^{}_r$. (e) A strong effect of shape anisotropy is observed when the field is applied along the increasing length of the sample, which is reflected in the very large value of remanent  magnetization $M^{}_r\approx 0.9$. (f) The dipolar interaction always decreases the coercive field $H^{}_c$. (g) For strongly interacting MNPs, $H^{}_c$ and $M^{}_r$ tend to zero when the field is applied normal to the plane of the sample. 

We have succeeded in obtaining well guidelines regarding how magnetic hysteresis properties may change in the dipolar interacting MNPs arrays. The obtained results contribute to expanding the fundamental comprehension of two-dimensional dipolar interacting system and offer exciting implications for the creation of self-assembled arrays of magnetic nanoparticles of desired magnetic response. The micromagnetic simulation showed that the dipolar interaction induces an effective ferromagnetic coupling in thin arrays (when the aspect ratio is very large). It means that one can modulate the magnetic properties of the ordered arrays of MNPs by varying the strength of shape anisotropy of the system. The latter can be changed by varying the width and length of the system. Our results are also relevant for the experimental samples, which can be divided into various categories depending on how they behave when an external magnetic field is applied to them.

\section*{ACKNOWLEDGMENTS}
Most of the numerical simulations presented in this work have been carried out in the Department of Physics, Indian Institute of Technology (IIT) Delhi. I am grateful to Prof. Varsha Banerjee for providing the computational facility at IIT Delhi. 

\bibliographystyle{h-physrev}
%\section*{References}
\bibliography{ref}
\newpage

\begin{table}[ht]
\begin{center}
\begin{tabular}{|l|l|}
 \hline
$a$ (nm) & $\Theta = D/KV$ \\ \hline
$10$&$1.75$  \\ \hline
 $12$&$1.01$   \\ \hline
$16$&$0.42$  \\ \hline
$20$&$0.22$    \\ \hline
$30$&$0.06$  \\ \hline
$40$&$0.03$\\ \hline
 \end{tabular}
 \end{center}
\caption{Evaluation of the ratio $\Theta = D/KV$ for Fe$_3$O$_4$ as a function of interparticle separation $a$  for cubical shaped particle. The lateral dimension of the particle is $l = 10$ nm.}

\label{table1}
\end{table}

\newpage

\newpage

\newpage

\newpage

\newpage
%\begin{figure}[!htb]
%\centering\includegraphics[scale=0.50]{Figure5.eps}
%\vspace{0.5cm}
%\caption{Dependence of magnetic hysteresis on aspect ratio $A^{}_r$ and dipolar interaction strength $\Theta$ when the field is applied along the longer axis of the sample, i.e., $\mu_o\vec{H}=H_o\hat{y}$. (a) $\Theta = 1.75$, (b) $\Theta = 1.01$, (c) $\Theta = 0.42$ and (d) $\Theta = 0.22$. The aspect ratio $A^{}_r$ is varied from 1.0 to 32.0 in each case. There is a strong dependence of $M^{}_r$ on the shape anisotropy introduced by increasing $A^{}_r$,  $M^{}_r$ reaches to $0.9$ when the field is applied along the alleger axis of the sample in the presence of strong dipolar interaction. The dipolar interaction decreases the coercive field $H^{}_c$ and $M^{}_r$. For weak dipolar interaction, i.e., $\Theta  = 0.22$, magnetic hysteresis follow the Stoner-Wohlfarth model ($H^{}_c \approx 0.48 H^{}_K$ and $M_r \approx 0.5$) irrespective of $A^{}_r$.}
%\label{figure5}
%\end{figure}

\newpage
\newpage

%\newpage
%\begin{figure}[!htb]
%\centering\includegraphics[scale=0.50]{Figure8.eps}
%\vspace{0.5cm}
%\caption{Remanent magnetization $M^{}_r$ variation as a function of dipolar interaction strength $\Theta$, aspect ratio $A^{}_r$ and direction of the applied field. (a) $\mu_o\vec{H}=H^{}_o\hat{x}$, (b) $\mu_o\vec{H}=H^{}_o\hat{y}$ and (c) $\mu_o\vec{H}=H^{}_o\hat{z}$. When the field is applied along the $x$-direction, $M^{}_r$ deceased with an increase in aspect ratio and dipolar interaction strength $\Theta$.  There is a strong shape anisotropy effect is observed when the field is applied along the longer axis of the sample. $M^{}_r$ increases with increase in $\Theta$ and $A^{}_r$, and it reaches to 0.9 with $A^{}_r=32.0$ and $\Theta = 1.75$. When the external field is applied normal to the plane of the sample $\mu_o\vec{H}=H_o\hat{z}$, $M^{}_r$ decreases with increase in $A^{}_r$ and $\Theta$, and it drops down to zero for $\Theta = 1.75$ and $A^{}_r = 32.0$.}
%\label{figure8}
%\end{figure}
\end{document}